\documentclass[aps,prl,twocolumn,floatfix,showpacs,superscriptaddress]{revtex4-2}

\usepackage{graphicx,amsmath,amsfonts,bm,mathtools}
\usepackage[colorlinks=true, citecolor=blue, linkcolor=blue, urlcolor=blue]{hyperref}
\usepackage[all]{hypcap}
\usepackage{amssymb}
\usepackage{mathrsfs}
\usepackage{clrscode3e,etoolbox}
\usepackage{varwidth}
\usepackage{newfloat}
\usepackage{dsfont}
\usepackage{color}
\usepackage{braket}
\usepackage[caption=false,subrefformat=parens,labelformat=empty]{subfig}
\DeclareCaptionType{Algorithm}
\usepackage[normalem]{ulem}
\usepackage{comment}
\usepackage[export]{adjustbox}

\usepackage[usenames,dvipsnames,svgnames,table]{xcolor}

\definecolor{carmine}{rgb}{0.59, 0.0, 0.09}

\DeclareMathOperator {\tr}{\text{Tr}}

\begin{document}

\title{Universal Characterization of Quantum Many-Body States through Local Information}
\author{Claudia Artiaco}
\thanks{These alphabetically ordered authors contributed equally.}
\affiliation{Department of Physics, KTH Royal Institute of Technology, Stockholm 106 91, Sweden}
\author{Thomas Klein Kvorning}
\thanks{These alphabetically ordered authors contributed equally.}
\affiliation{Department of Physics, KTH Royal Institute of Technology, Stockholm 106 91, Sweden}
\author{David Aceituno Chávez}
\affiliation{Department of Physics, KTH Royal Institute of Technology, Stockholm 106 91, Sweden}
\author{Loïc Herviou}
\affiliation{Institute of Physics, Ecole Polytechnique Fédérale de Lausanne (EPFL), CH-1015 Lausanne, Switzerland}
\affiliation{Université Grenoble Alpes, CNRS, LPMMC, 38000 Grenoble, France}
\author{Jens H. Bardarson}
\affiliation{Department of Physics, KTH Royal Institute of Technology, Stockholm 106 91, Sweden}

\begin{abstract}
We propose a universal framework for classifying quantum states based on their scale-resolved correlation structure.
Using the recently introduced information lattice, which provides an operational definition of the total amount of correlations at each scale, we define intrinsic characteristic length scales of quantum states.
We analyze ground and midspectrum eigenstates of the disordered interacting Kitaev chain, showing that our framework provides a novel unbiased approach to quantum matter.
\end{abstract}

\maketitle

Quantum information theory has emerged as a powerful tool for understanding the complex behavior of many-body systems~\cite{zeng2019quantum,laflorencie2016quantum,eisert2010colloquium,srednicki1993entropy,hastings2007area,wolf2008area,vidal2003entanglement,calabrese2005evolution,kitaev2006topological,levin2006detecting}.
In condensed matter, classifying states by the scaling of the von Neumann entropy in subsystems is a standard approach~\cite{laflorencie2016quantum,eisert2010colloquium}.
The ground states of local gapped Hamiltonians follow the area law~\cite{srednicki1993entropy,hastings2007area,wolf2008area}, and in one dimension are effectively captured by matrix product states, underlying the success of the density-matrix renormalization group~\cite{white1992density,ostlund1995thermodynamic,schollwock2011density}.
Critical ground states show logarithmic violations of the area law with universal prefactors~\cite{vidal2003entanglement,calabrese2005evolution}.
Midspectrum states of ergodic Hamiltonians are highly entangled following a volume law that, according to the eigenstate thermalization hypothesis~\cite{deutsch1991quantum,srednicki1994chaos,rigol2008thermalization}, gives rise to thermalization of subsystems.
In two dimensions, topological order is characterized by a universal constant contribution to the von Neumann entropy~\cite{kitaev2006topological,levin2006detecting}.
Additionally, advancements in understanding the interplay between gravity and quantum mechanics often leverage quantum information theory, particularly through black hole entropy~\cite{bekenstein1973black,hawking1975particle} and the Ryu-Takayanagi formula~\cite{ryu2006holographic}.
It is even proposed that spacetime may emerge from information structures~\cite{vanraamsdonk2010building,swingle2012entanglement,cao2017space,cao2018bulk}.

\begin{figure}[t]
    \raggedright
    \begin{minipage}{31.0mm}%
    \subfloat{\includegraphics[keepaspectratio,width=0.92\linewidth,valign=b]{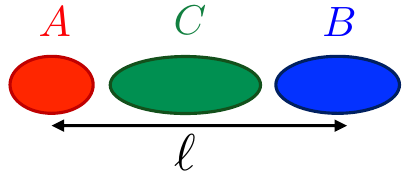}%
        \llap{\parbox[b]{0.5mm}{(a)\\\rule{0ex}{7.5mm}}}%
        \label{fig:schematic_sets}}\\
        
        \vspace{-3mm}\subfloat{\includegraphics[keepaspectratio,width=\linewidth,valign=b]{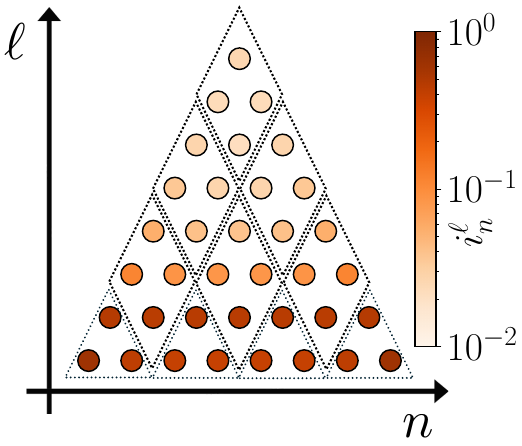}%
        \llap{\parbox[b]{1.5mm}{(b)\\\rule{0ex}{21.5mm}}}%
        \label{fig:new_unit_cell}}%
    \end{minipage}
    \hspace{1mm}
    \begin{minipage}{52.0mm}
        \subfloat{\includegraphics[keepaspectratio, width=\linewidth,valign=b]{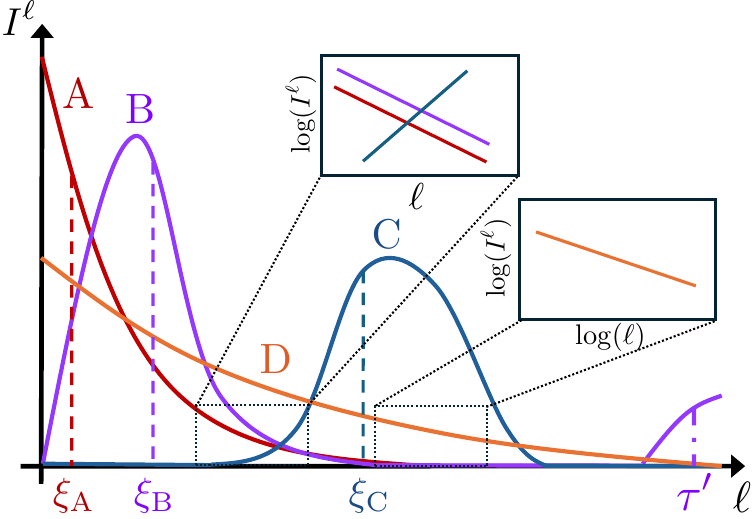}%
        \llap{\parbox[b]{10mm}{(c)\\\rule{0ex}{34mm}}}%
        \label{fig:schematic_infor_per_scale}}%
    \end{minipage}%
    \vspace{-2mm}%
    \caption{(a) Schematic of subregions  $A$, $B$ and $C$ in a quantum system.
    $A$ and $B$ are disjoint and at distance $\ell$. 
    (b) Information lattice for a chain of 8 physical sites in the critical ground state of the clean noninteracting Kitaev chain [Eq.\eqref{eq:hamiltonian}].
    Information lattice sites (black circles) are labeled by $(n, \ell)$ and associated with local information $i^{\ell}_{n}$ [Eq.~\eqref{eq:local_information}], which is quantified by the intensity of the color.
    Dotted areas illustrate the information lattice after redefining physical sites as blocks of $\ell^*=2$ sites [discussion surrounding Eq.~\eqref{eq:critical-information}].
    (c) The information per scale $I^\ell$ for four prototypical examples: (A) localized state peaked at $\ell=0$; (B) localized states peaked at finite $\ell$ and with an $\mathcal{O}(1)$ correction of information at system-size scales; (C) fully ergodic state; (D) critical scale-invariant state. The expected correlation lengths are indicated as $\xi$ [Eq.~\eqref{eq:xidef}] and $\tau^\prime = L - 1 - \tau$ [Eq.~\eqref{eq:taudef}]. 
    Insets: Enlargement of the dotted areas with rescaled axes to show the scaling behaviors.
    }
    \label{fig:schematic}
\end{figure}

These examples motivate the study of the local structure of quantum information in many-body states. 
Approaches that go beyond the binary area and volume law classification include various multipartite entanglement measures~\cite{doherty2005detecting,szalay2015multipartite,walter2016multipartite}, the use of quantum Fisher information~\cite{pezze2009entanglement,hyllus2012fisher,toth2012multipartite,strobel2014fisher,hauke2016measuring}, and entanglement link representations~\cite{singha2020entanglement,singha2021Link,santalla2023entanglement}.
The fundamental challenge is to determine precisely on which scales information in different parts of a system resides.
A first attempt might consider correlation functions between local operators at distance $\ell$.
While the connected correlation function $\langle O_A O_B \rangle - \langle O_A \rangle \langle O_B \rangle$---where $O_A$ is an operator acting in $A$ and $O_B$ in $B$ with $A$ and $B$ disjoint subregions at distance $\ell$, as sketched in Fig.~\ref{fig:schematic_sets}---is a natural candidate, it does not provide a complete and unbiased measure of the \textit{total} amount of correlations at scale $\ell$.
To achieve this, one would in principle need to measure correlation functions for all possible local operators.
When multiple characteristic length scales are present in the system, different correlation functions may yield different results~\cite{colbois2024interaction}.
The mutual information instead gives a precise meaning to the total correlations between $A$ and $B$, as it provides an upper bound to normalized connected correlation functions~\cite{groisman2005quantum,wolf2008area}.
However, it exactly captures the total correlations on scale $\ell$ only if the state is in a product state between $A \cup B$ and the rest of the system.
Nonzero mutual information between disjoint regions may be due to both $A$ and $B$ being correlated with $C$ (see Fig.~\ref{fig:schematic_sets}) and thus not necessarily directly correlated.
The mutual information can also miss information on scale $\ell$.
For example, the state can differ from an infinite temperature state on $A \cup C \cup B$, yet tracing out one of $A, B$ or $C$ leaves a maximally mixed density matrix that has zero information~\footnote{For instance, the density matrix in $A \cup C \cup B$ could be $2^{-\ell-1}(\mathds{1} + \sigma_z^{\otimes (\ell+1)})$.}.

What then are the requirements for a well-defined notion of information at scale $\ell$?
First, it should decompose the total information in a region according to the corresponding scales, and reduce to the mutual information when it correctly captures the total correlations at scale $\ell$.
It should also be local: information in $A$ should not be affected by a local unitary acting outside of $A$, and a local unitary on nearest neighbors should only be able to move information from scale $\ell$ to $\ell \pm 1$.
Information appears intrinsically nonlocal and it is thus \textit{a priori} not obvious that one can construct such a local decomposition to treat information as a locally conserved quantity.
In this Letter, we show that the ``information lattice," which in Refs.~\cite{klein2022time,artiaco2024efficient,harkins2025nanoscale} was applied to perform efficient time evolution of large-scale quantum states, provides precisely such a decomposition for one-dimensional systems that can be used to fully characterize states in different universality classes.
We analyze the features of the information per scale in various example states encountered in quantum matter: localized, critical and ergodic states, as well as trivial and topological states, as summarized in Fig.~\ref{fig:schematic_infor_per_scale}.
The scale decomposition of total correlations obtained via the information lattice provides uniquely unbiased and universal definitions of characteristic length scales for generic quantum states.

\textit{Scales and subsystems}---We consider a one-dimensional system composed of $L$ sites, each representing a quantum degree of freedom with Hilbert space dimension $d$.
Following Refs.~\cite{klein2022time,artiaco2024efficient}, we define the subsystem $\mathcal{C}_n^{\ell}$ as the set of $\ell + 1$ contiguous physical sites centered around $n$.
Thus, when $\ell$ is odd $n$ is a half-integer, and when $\ell$ is even $n$ is an integer.
We refer to $\ell$ as \textit{scale}.
We define $\bar{\mathcal{C}}_n^{\ell}$ as the complement of $\mathcal{C}_n^{\ell}$.
The state of the entire system is given by the density matrix $\rho$ with dimension $\dim(\rho) = d^{L}$.
The subsystem density matrix is $\rho_n^{\ell} = \tr_{\bar{\mathcal{C}}_n^{\ell}}(\rho)$, where $\tr_{\bar{\mathcal{C}}_n^{\ell}}$ is the trace operator over the complement $\bar{\mathcal{C}}_n^{\ell}$.

\textit{Information lattice}---The von Neumann information $I(\rho)$ quantifies the total information in the quantum state $\rho$~\cite{von2013mathematische,vonNeumann1927thermodynamik}.  
$I(\rho)$ is the average number of bits that can be predicted about measurement outcomes from $\rho$ and equals the deficit of the von Neumann entropy $S(\rho)$ from its maximum value, $I(\rho) = \log_2[\dim(\rho)] - S(\rho) = \log_2[\dim(\rho)] + \tr[\rho \log_2(\rho)]$.
For a pure state ($\rho^2 = \rho$), $I(\rho) = L \log_{2}(d)$.
The von Neumann information of a subsystem density matrix $I(\rho^\ell_n)$ analogously quantifies the information concerning observables acting only in $\mathcal{C}^\ell_n$. 
For a state $\rho$, we define the local information in subsystem $\mathcal{C}_n^{\ell}$ on scale $\ell$ as
\begin{eqnarray}
\label{eq:local_information}
i^\ell_n = I(\rho^\ell_n) - I(\rho^{\ell-1}_{n-1/2}) - I(\rho^{\ell-1}_{n+1/2}) + I(\rho^{\ell-2}_{n}),
\end{eqnarray}
where it is implicit that the von Neumann information of empty subsystems is zero. 
The local information $i^\ell_n$ is the information in $\rho^\ell_n$ that cannot be obtained from smaller subsystem density matrices.
It quantifies the total amount of correlations present at scale $\ell$ in the region centered around $n$, which can be accessed by measuring correlation functions for all possible local operators within that spatial region.
The information lattice is the two-dimensional triangular structure shown in Fig.~\ref{fig:new_unit_cell}, where the lattice sites $(n, \ell)$ are uniquely labeled by the spatial location $n$ (horizontal axis) and scale $\ell$ (vertical axis). 
Given a state $\rho$, we associate each site with $i^{\ell}_{n}$.
Fig.~\ref{fig:new_unit_cell} illustrates the information lattice for the critical scale-invariant ground state of the clean noninteracting Kitaev chain in Eq.~\eqref{eq:hamiltonian} below.
While we focus on pure states, the information lattice is also well defined for mixed states~\cite{klein2022time,artiaco2024efficient,harkins2025nanoscale}.

Local information on different scales and different subsystems is independent, meaning that the sum of local information in all lower-scale subsystems equals the total subsystem information~\cite{klein2022time}, $ I(\rho^\ell_n) = \sum_{(n^\prime,\ell^\prime) \in S^\ell_n} i^{\ell^\prime}_{n^\prime}$, with $S^\ell_n = \{ (n',\ell') \, | \, \mathcal{C}^{\ell'}_{n'} \subseteq \mathcal{C}^{\ell}_{n} \}$.
Thus, the local information is a decomposition of the total information into location and scale: $I(\rho) = \sum_{\mathrm{all} \; (n,\ell)} i^\ell_n$.
The information per scale,
\begin{equation}
    I^\ell = \sum_n i^\ell_n,
    \label{eq:info_per_scale}
\end{equation}
which quantifies the total correlations on scale $\ell$ over the entire state, provides unbiased and universal definitions of correlation lengths.

\textit{Characterizing states via the information lattice}---%
We begin by considering localized states (curves A and B in Fig.~\ref{fig:schematic_infor_per_scale}) where all the information, except for $\mathcal{O}(1)$ corrections, is present on scales $\ell \ll L$.
Example A is a localized state close to a local product state, showing a peak of the information per scale at $\ell=0$.
A localized state can also be close to a product state of singlets with short-range correlations, as in example B; such a state shows a peak at finite $\ell \sim \mathcal{O}(1)$ indicating that lower-scale subsystems are mixed and correlations are primarily concentrated between the singlet pairs.
To capture these features, we introduce the \textit{expected correlation length}
\begin{align}
    \xi = \frac{\sum_{\ell=0}^{\lfloor L/2\rfloor} \ell \, I^\ell}{\sum_{\ell=0}^{\lfloor L/2\rfloor}I^\ell},
    \label{eq:xidef}
\end{align}
which quantifies the scale at which correlations are most likely to occur within the state; we sum only to $\lfloor L/2\rfloor$ to avoid the $\mathcal{O}(1)$ corrections at system-size scales discussed below.
The length $\xi$ is represented by the vertical dashed lines in Fig.~\ref{fig:schematic_infor_per_scale}.

Localized states are also characterized by the exponential decay of $I^\ell$ across scales as it diminishes away from the short-scale maximum~\cite{brandao2013area} (see inset). 
We denote the \textit{correlation decay length} of localized states as $\lambda$ and define it using a least-squares fit over the range $[\lfloor L/4\rfloor, \lceil L/2\rceil]$~%
\footnote{The lower limit is set as large as possible to capture the exponential decay of information at large scales while ensuring enough data points for fitting.
The upper limit prevents contributions from the $\mathcal{O}(1)$ correction at system-size scales.}:
\begin{align}
    \ln(I^\ell) &\sim -\ell/\lambda + \text{const.}, & \ell \in [\lfloor L/4\rfloor, \lceil L/2\rceil].
    \label{eq:lambdadef}
\end{align}

In addition to the information at small scales, localized states can have an $\mathcal{O}(1)$ correction at scales comparable to the system size (curve B in Fig.~\ref{fig:schematic_infor_per_scale}).
This correction could have different origins.
One possibility is a cat state superposition of $n$ states that has $\log_2(n)$ bits of information at system-size scales, as only observables acting on $\mathcal{O}(L)$ number of sites can access the relative phases between the states.
Another possibility is the presence of accidental edge correlations; for instance, a singlet shared between the edges would contribute two bits at the largest scale.
Finally, such an $\mathcal{O}(1)$ correction can be due to topology.
Fermionic noninteracting topological phases are characterized by $N$ occupied edge modes in the filled Bogoliubov-de Gennes band~\cite{kitaev2001unpaired,chiu2016Classification}.
There are then $N$ independent binary questions that one can answer regarding correlations between the edges~\footnote{Notice that, in contrast to a cat-state superposition, the observables to access the information concerning topological edge correlations act on $\mathcal{O}(1)$ number of sites that are spread across an $\mathcal{O}(L)$-sized region.}.
This implies that the total information on large scales $\Gamma$ is
\begin{align}
    \Gamma = \sum_{\ell=\lfloor L/2 \rfloor}^{L-1} I^\ell = N,
\end{align}
up to exponentially small corrections.
Two states belong to the same symmetry-protected topological phase if they can be connected by a symmetry-preserving local unitary circuit with finite depth~\cite{chen2010local,chen2011classification}.
$\Gamma$ is invariant under the application of such a local unitary circuit and therefore serves as a universal characteristic of topological phases.
While our discussion focused on fermionic states, the analysis is general and equally applicable to other systems, such as Aﬄeck-Kennedy-Lieb-Tasaki~\cite{affleck1987rigorous} and Haldane chains~\cite{haldane1983nonlinear}.

Edge correlations also have associated length scales.
In analogy with $\xi$, we define the ``expected edge-correlation length" as~\footnote{Similarly to Eqs.~\eqref{eq:xidef} and \eqref{eq:lambdadef}, the limits are set to avoid nontopological contributions.}
\begin{align}
\label{eq:taudef}
    \tau = L - 1 - \frac{\sum_{\ell=\lfloor L/2\rfloor}^{L-1} \ell\, I^\ell}{\sum_{\ell=\lfloor L/2\rfloor}^{L-1} I^\ell},
\end{align}
which is represented by the dashed-dotted purple line in Fig.~\ref{fig:schematic_infor_per_scale}. 
An edge-correlation decay length measuring the localization length of (topological) edge correlations can also be defined, similarly to Eq.~\eqref{eq:lambdadef}.

Fully ergodic states (curve C in Fig.~\ref{fig:schematic_infor_per_scale}), such as midspectrum eigenstates of interacting local Hamiltonians, exhibit an extensive amount of information at half the system size.
$\xi$ in \eqref{eq:xidef} is then extensive, and 
$\lambda$ in \eqref{eq:lambdadef} is negative and quantifies the decay away from half the system size.
Its value can be derived from statistical considerations:
up to $L/2$ it takes 4 times as many parameters to encode information at scale $\ell$ compared to $\ell - 1$, as the reduced density matrix at scale $\ell$ contains 4 times as many elements.
We then expect $I^{\ell}$ to increase by a factor of $4$ each time $\ell$ increases by $1$, giving $\lambda = -\ln(4)$.
This argument is verified exactly by Haar-random states~\cite{page1993average}.
Different scaling behaviors might appear, for instance, for an eigenstate in a Krylov subspace where the Hamiltonian follows the Krylov-restricted eigenstate-thermalization hypothesis~\cite{bernevig2021}.
Ergodic states at finite temperature, such as high-energy eigenstates away from midspectrum, present both a peak of information at small scales and at half the system size.
The definitions of $\xi$ and $\lambda$ can be adapted to characterize both behaviors.

Finally, we focus on critical states (curve D in Fig.~\ref{fig:schematic_infor_per_scale}).
The hallmark of criticality is long-range scale invariance. 
In a lattice theory, scale invariance means that the physics remains unchanged when single sites are rescaled as blocks of sites.
Given the two-dimensional structure of the information lattice, redefining single sites as blocks of $\ell^*$ sites transforms the local information $i^\ell_n$ into the sum of $(\ell^*)^2$ original values (see Fig.~\ref{fig:new_unit_cell}).
Simultaneously, the length scale renormalizes as $\ell \rightarrow \ell / \ell^*$ and the average local information on scale $\ell$, $i^\ell=\langle i^\ell_n\rangle$, changes to $i^\ell \rightarrow (\ell^*)^2 i^{\ell / \ell^*}$.
The information distribution invariant under such a transformation is
\begin{align}
    i^{\ell} = \frac{\alpha}{\ell^2},
    \label{eq:critical-information}
\end{align}
where $\alpha$ is a constant dependent on the scale-invariant state. 
In a finite-size system, the behavior \eqref{eq:critical-information} is expected to hold only at intermediate scales; at large scales, finite-size effects dominate and the behavior takes a boundary-condition-dependent form $i^{\ell} = f(\ell/L)$.

\begin{figure}[t]
    \raggedright
    \begin{minipage}{86.4mm}%
        \subfloat{\includegraphics[keepaspectratio,width=0.498\linewidth,valign=b]{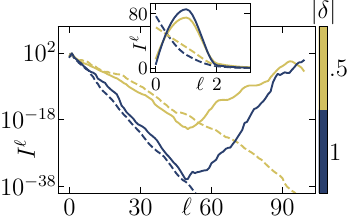}%
        \llap{\parbox[b]{18.5mm}{(a)\\\rule{0ex}{18.4mm}}}%
            \label{fig:Gaussian}}%
            \hspace{0.3mm}
        \subfloat{\includegraphics[keepaspectratio,width=0.485\linewidth,valign=b]{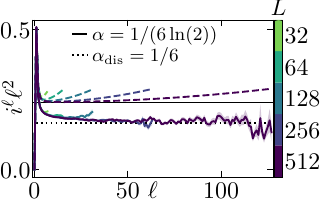}%
        \llap{\parbox[b]{18.5mm}{(b)\\\rule{0ex}{18.4mm}}}%
        \label{fig:critical}}%
    \end{minipage}

    \vspace{-2mm}%
    \caption{(a) The information per scale $I^\ell$ for the ground states of the noninteracting ($g=0$) Kitaev chain \eqref{eq:hamiltonian} at $\delta=\pm 0.5, \pm1$ (dashed lines for $\delta<0$; solid lines for $\delta>0$) for $L=100$ sites. 
    We find: for $\delta=-0.5$, $\xi=0.60$, $\lambda=0.93$; for $\delta=-1$, $\xi=0.29$, $\lambda=0.54$; for $\delta=0.5$, $\xi=1.0$, $\lambda=0.82$, $\tau=5.5$; for $\delta=1$, $\xi=1.0$, $\lambda=0.56$, $\tau=0.64$.
    Inset: Short-scale zoom-in with lines showing the monotonic cubic interpolation of the data.
    Single disorder realizations are shown.
    (b) $i^\ell$ for intermediate $\ell$'s follows the scale-invariant behavior \eqref{eq:critical-information} in the ground states of both clean ($t_{2j-1} = t_{2j} = \mathrm{const.}$) (dashed) and disordered ($\delta = 0$) (solid) noninteracting Kitaev chains at the critical point. 
    Solid lines are averages over more than 40k disorder realizations and the width of the shaded areas is the standard error.
    }
    \label{fig:ground-states}
\end{figure}

\textit{Ground states of the (disordered) Kitaev chain}---%
We demonstrate the use of the information lattice by analyzing the eigenstates of the disordered interacting Kitaev Hamiltonian on $L$ sites with open boundary conditions
\begin{equation}
\label{eq:hamiltonian}
    H=-i\sum_{j=1}^{2L-1} t_j\gamma_j \gamma_{j+1} + g\sum_{j=1}^{2L-3}\gamma_j \gamma_{j+1} \gamma_{j+2} \gamma_{j+3},
\end{equation}
where $\gamma_{2j-1}=c_j + c^\dagger_j$ and $\gamma_{2j}=i(c_j - c^\dagger_j)$ are Majorana operators expressed in terms of fermion creation ($c_j^\dagger$) and annihilation ($c_j$) operators on site $j$. 
The parameters $t_{j}$ are uniformly distributed in the intervals $t_{2j-1} \in [0, e^{-\delta/2}]$ and $t_{2j} \in [0, e^{\delta/2}]$; $g$ is the interaction strength. 
This model has recently been the subject of intense research~\cite{lobos2012interplay,crepin2014nonperturbative,kjall2014many,milsted2015Statistical,gergs2016topological,mcginley2017robustness,monthus2018topological,venderley2018machine,karcher2019disorder,moudgalya2020perturbative,sahay2021emergent,roberts2021infinite,wahl2022local,laflorencie2022topological,chepiga2023topological}.

By using the information per scale $ I^\ell $, we characterize the ground states of the noninteracting ($g=0$) Hamiltonian~\eqref{eq:hamiltonian} for $\delta = \pm 0.5, \pm 1$, as depicted in Fig.~\ref{fig:Gaussian}.
Our results show that the correlation decay length $\lambda$ remains approximately the same within pairs of disorder realizations sharing the same $|\delta|$.
This is due to the duality of the Hamiltonian under the transformation $\delta \rightarrow -\delta$ via $\gamma_j \rightarrow \gamma_{j+1}$, which guarantees that the distribution of correlation decay lengths is preserved when the sign of $\delta$ is reversed.
The inset of Fig.~\ref{fig:Gaussian} illustrates that the expected correlation length $\xi$ significantly varies between realizations with opposite signs of $\delta$.
This is also accounted for by the duality: an on-site product state for negative sign of $\delta$ is mapped to a ``dimer'' state for positive $\delta$, determining two different $I^{\ell}$ distributions akin to the ones shown in Fig.~\ref{fig:schematic_infor_per_scale} (A and B, respectively).
For the single disorder realizations in Fig.~\ref{fig:Gaussian} we find $\lambda \approx 1$ at $\delta = \pm 0.5$, $\xi \approx 0.6$  at $\delta = - 0.5$ (indicating that a substantial amount of the information is on site) and $\xi \approx 1$ at $\delta = 0.5$ (indicating that most correlations occur between nearest neighbors).

A key difference between the data for different signs of $\delta$ in Fig.~\ref{fig:Gaussian} is the behavior of $I^\ell$ at system-size scales, which in this model is due to topological correlations between the edges.
The eigenstates with $\delta < 0$ have $\Gamma \approx 0$, whereas those with $\delta > 0$ have $\Gamma \approx 1$, consistent with expectations~\cite{laflorencie2022topological}.
For $\delta=0.5$, the expected edge-correlation length is $\tau \approx 5.5$. 
This deviation from $\tau \approx 0$, which would signal correlations precisely between the two edge physical sites, indicates that, for small positive $\delta$, the broad distributions of $t_j$'s can drive regions near the edges into the trivial topological phase, shifting edge correlations inward, away from the boundaries.

Finally, we consider the ground states of the noninteracting Kitaev model at the transition between the trivial and topological phases.
The clean model with $ t_{2j-1} = t_{2j} = \mathrm{const.}$ is described at low energy by a $(1+1)$-dimensional conformal field theory~\cite{calabrese2004entanglement,difrancesco2006conformal}, which predicts $ \alpha = c/(3\ln2)$ [see Eq.~\eqref{eq:critical-information}], where $c = 1/2$ is the central charge, as reproduced in Fig.~\ref{fig:critical}.
For the disordered Hamiltonian~\eqref{eq:hamiltonian}, $ \delta = 0 $ also corresponds to the trivial-to-topological phase transition where the ground state is scale-invariant~\cite{fisher1992random,fisher1995critical}.
Analytical predictions~\cite{refael2004entanglement,laflorencie2005scaling,refael2009criticality} give $ \alpha_\mathrm{dis} = 1/6 $.
This behavior is captured in our numerics at scales $ \ell \lesssim L/4 $.
To reduce finite-size effects, also reported in other studies~\cite{laflorencie2005scaling,chepiga2024resilient}, we compute $i^\ell$ by averaging local information $i^\ell_n$ over the central sites of the information lattice within an equilateral triangle of base $L/4$ at $\ell=0$.

\begin{figure}[t]
    \raggedright
    \begin{minipage}{86.4mm}%
        \subfloat{\includegraphics[keepaspectratio,width=0.48\linewidth,valign=b]{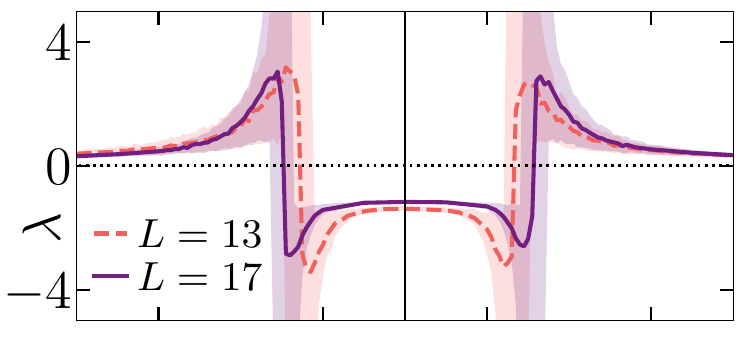}%
        \llap{\parbox[b]{7mm}{(a)\\\rule{0ex}{12.3mm}}}%
            \label{fig:lambda}}%
        \subfloat{\includegraphics[keepaspectratio,width=0.48\linewidth,valign=b]{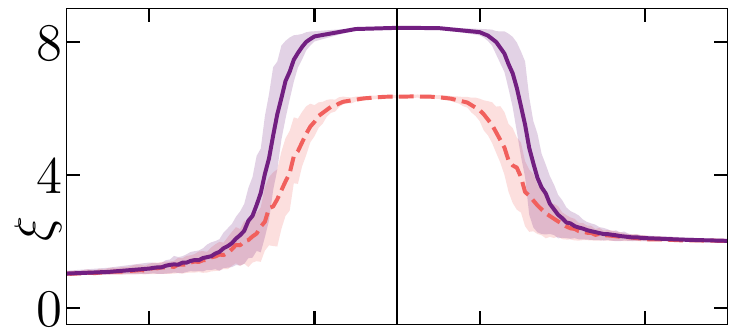}%
        \llap{\parbox[b]{7mm}{(b)\\\rule{0ex}{12.3mm}}}%
        \label{fig:xi}}%
        \vspace{-4mm}%
    \end{minipage}
    \begin{minipage}{86.4mm}%
        \subfloat{\includegraphics[keepaspectratio,width=0.48\linewidth,valign=b]{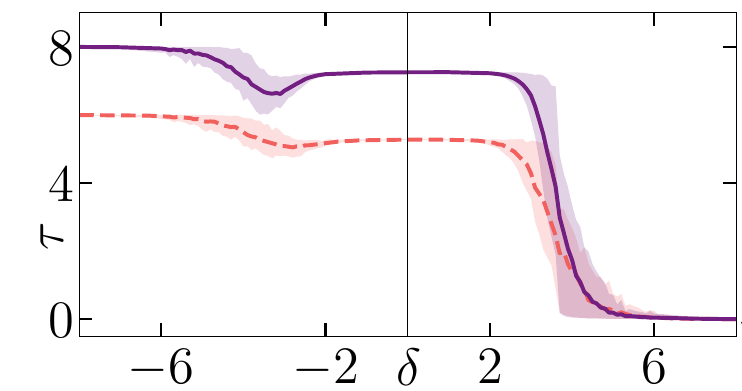}%
        \llap{\parbox[b]{7mm}{(c)\\\rule{0ex}{15.2mm}}}%
            \label{fig:tau}}%
        \subfloat{\includegraphics[keepaspectratio,width=0.48\linewidth,valign=b]{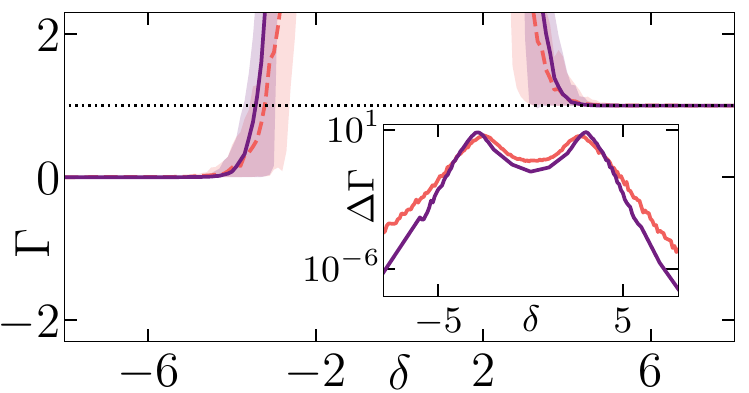}%
        \llap{\parbox[b]{7mm}{(d)\\\rule{0ex}{15.2mm}}}%
        \label{fig:Gamma}}%
    \end{minipage}
    \vspace{-2mm}%
    \caption{Midspectrum even-parity eigenstates of model \eqref{eq:hamiltonian} as a function of $\delta$, for $g=0.5$, $L=13,17$.
    The lines are the distribution medians and the shaded areas are the narrowest intervals encompassing 75\% of the disorder realizations (more than 100 for each data point). 
    (a) The correlation decay length $\lambda$.
    The dotted horizontal line marks the value 0.
    (b) The expected correlation scale $\xi$.
    (c) The expected edge-correlation scale $\tau$.
    (d) The information on scales larger than half the system size $\Gamma$.
    The dotted horizontal line marks the value 1.
    The inset shows the width of the narrowest interval encompassing 75\% of the realizations, $\Delta\Gamma$.
    }
    \label{fig:mbl}
\end{figure}

\textit{Midspectrum states of the disordered interacting Kitaev chain}---%
We now examine the midspectrum eigenstates of the interacting Kitaev Hamiltonian~\eqref{eq:hamiltonian} for $g=0.5$, system sizes $L=13,17$, and disorder values $\delta \in \left[ -8, 8 \right]$.
For each disorder realization, we consider the even fermion parity eigenstate with energy eigenvalue closest to zero.
Given the expected phase diagram for this model~\cite{laflorencie2022topological}, we anticipate that typical eigenstates are either localized and topologically trivial ($\delta \lesssim - \delta_c$), ergodic ($- \delta_c \lesssim \delta \lesssim \delta_c$) or localized and topological ($\delta \gtrsim \delta_c$), where $\delta_c > 0$.  
The expected correlation length $\xi$ is $\mathcal{O}(1)$ for localized states and scales with system size for fully ergodic states.
The expected edge-correlation length $\tau$, instead, is $\mathcal{O}(1)$ for localized states with edge correlations and becomes extensive for ergodic or localized states without edge correlations.
Figure~\ref{fig:mbl} shows properties of the distributions of the correlation decay length $\lambda$, the expected correlation length $\xi$, the expected edge-correlation length $\tau$, and the total information at scales larger than half the system size $\Gamma$.
Up to finite-size corrections, the duality $ \delta \rightarrow -\delta$ is still valid. 
The correlation decay length for fully ergodic states is approximately $-\ln(4)$ (see Fig.~\ref{fig:lambda}), as predicted, and the scaling of $\xi$ reliably distinguishes between the three phases (see Fig.~\ref{fig:xi}).
Fig.~\ref{fig:tau} shows that $\tau$ correctly captures the onset of the topological phase, becoming independent of system size at approximately the same $\delta$ as $\xi$, which also coincides with the point where $\lambda$ turns positive.
At the same disorder strength, $\Gamma$ quickly converges to 1, as shown in Fig.~\ref{fig:Gamma}.

\textit{Discussion}---%
We demonstrated that the local information $i^\ell_n$ provides \textit{the} operational meaning of information at scale $\ell$ within a system.
This derives from its definition as the information in a subsystem of extent $\ell$ that is absent in any smaller subsystems.
Summing over $n$ provides an unbiased measure of the information at scale $\ell$ that we used for characterizing states and defining characteristic length scales.
These correlation lengths are intrinsic to the many-body state distinguishing them from other length scales that depend on the parent Hamiltonian.
For instance, Anderson localization lengths of single-particle orbitals~\cite{evers2008anderson,kramer1993localization} and $l$-bit localization lengths in many-body localization~\cite{alet2018many,abanin2019colloquium,sierant2025many} cannot be accessed from a single many-body state.

We demonstrated the usefulness of the local-information approach by characterizing finite-size states from numerical simulations.
However, our framework holds broader conceptual significance, offering a new lens to understand quantum matter.
This is similar to the role of other quantum information tools in condensed matter physics: while entanglement entropy does not reveal anything beyond what relevant observables could provide, it abstracts away specific observables and highlights universal properties---like topology---independent of the particular physics at hand.
Local information acts as a similar filter by placing scale at center stage and allowing for precise quantification of the spatial structure of correlations.
Scale is essential to understand processes such as decoherence and thermalization where information moves to inaccessible large-scale degrees of freedom~\cite{klein2022time,artiaco2024efficient,harkins2025nanoscale}, and to determine the size of a probe needed to sense physical phenomena.
The small-scale section of the information lattice is in principle experimentally accessible through quantum state tomography~\cite{smithey1993measurement,leibfried1996experimental,leonhardt1995ulf,cramer2010efficient,lvovsky2009continuous,baumgratz2013scalable} or entropy measurements~\cite{islam2015measuring,vermersch2024many}, without any a priori knowledge of the state or relevant observables, guiding the reconstruction of the most salient features of a state.
Furthermore, while in this article we have focused on characterizing pure states, the information lattice is directly applicable to mixed states and can be used analogously to characterize mixed-state phases and phase transitions~\cite{coser2019classification,rakovszky2024defining,sang2024mixed,sala2024spontaneous,sang2025stability,ma2025symmetry}.
The generality of this framework allows its wide applicability.
By providing a universal description of quantum states, it is not only relevant in condensed matter but also in any other field that requires an operational definition of local information.

\textit{Acknowledgements}---%
We thank Miguel F. Martínez for useful discussions. This work received funding from the European Research Council (ERC) under the European Union’s Horizon 2020 research and innovation program (Grant Agreement No.~101001902) and the Knut and Alice Wallenberg Foundation (KAW) via the project Dynamic Quantum Matter (2019.0068). T.~K.~K. acknowledges funding from the Wenner-Gren Foundations. The computations were enabled by resources provided by the National Academic Infrastructure for Supercomputing in Sweden (NAISS), partially funded by the Swedish Research Council through grant agreement no. 2022-06725.

\begin{figure}[t]
    \centering
    \includegraphics[keepaspectratio,width=\linewidth,valign=b]{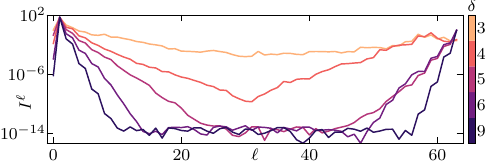}
    \vspace{-2mm}
    \caption{Information per scale $I^\ell$ for the ground states of the interacting ($g=0.5$) model \eqref{eq:hamiltonian} obtained with the density-matrix renormalization group at $\delta=3,4,5,6,9$ for $L=64$.
    Single disorder realizations are shown.
    }
    \label{fig:DMRG}
\end{figure}

\textit{End matter}---The information lattice can be computed for large systems encoded as matrix product states (MPS), as demonstrated in Fig.~\ref{fig:DMRG} for ground states of the disordered interacting Kitaev chain at $g=0.5$ for $L=64$ sites obtained via the density-matrix renormalization group algorithm~\cite{aceituno2024thermalization}.
Computing the information lattice requires calculating the von Neumann entropy $S(\rho^\ell_n)$ for every contiguous subsystem $\mathcal{C}^\ell_n$, which is given by either a single-cut partition ($\mathcal{C}^\ell_n|\bar{\mathcal{C}}_n^{\ell}$) or a double-cut partition ($\bar{\mathcal{C}}_n^{\ell} | \mathcal{C}^\ell_n | \bar{\mathcal{C}}_n^{\ell}$) of the physical chain.
For single-cut partitions, the entropy is obtained “for free” in the MPS formalism as $S(\rho^\ell_n) = -\sum_{k=1}^\chi s_k^2 \log_2 s_k^2$, where $s_k$ are the singular values in the diagonal bond matrix connecting $\mathcal{C}^\ell_n$ to its complement $\bar{\mathcal{C}}_n^{\ell}$.
Entropies for double-cut partitions can be obtained by $S(\rho^\ell_n) = -\sum_k \lambda_k \log_2 \lambda_k$ where $\lambda_k$ are the eigenvalues of either the subsystem density matrix $\rho^\ell_n$ or the transfer matrix $T^\ell_n$~\footnote{Following Ref.~\cite{svetlichnyy2024matrix}, we define $T^\ell_n = \sum_{\boldsymbol{\sigma}} (M^{\boldsymbol{\sigma}})^\dagger M^{\boldsymbol{\sigma}}$, where $M^{\boldsymbol{\sigma}}$ is a left-and-right normalized MPS for the degrees of freedom $\boldsymbol {\sigma} = \sigma_i\dots\sigma_j$ in the subsystem $\mathcal{C}^\ell_n$. In Vidal's notation~\cite{Vidal2003efficient}, $M^{\boldsymbol{\sigma}} =
\Lambda^{[i-1]} \Gamma^{[i]\sigma_i} \dots \Gamma^{[j]\sigma_j} \Lambda^{[j]}$. The size of $T^\ell_n$ depends solely on the bond dimensions of $M$.
}.
For pure states ($\rho^2=\rho$) one can exploit the relation $S(\rho^\ell_n)=S(\bar{\rho}^\ell_n)$, where $\bar{\rho}^\ell_n = \tr_{\mathcal{C}^\ell_n}(\rho)$ is the density matrix of the complement subsystem, which is smaller than $\rho^\ell_n$ when $\ell \geq \lfloor L/2 \rfloor$. 
This relationship facilitates the calculation of entries at the top of the information lattice.

The numerical cost of contracting and calculating the eigenvalues of $\rho^\ell_n$, $\bar{\rho}^\ell_n$ or $T^\ell_n$ may vary greatly for any given subsystem. 
For example, $\rho^\ell_n$ grows exponentially with subsystem size as $4^{\ell+1}$, while $T^\ell_n$ scales as $(\chi_L \chi_R)^2$, where $\chi_L$ and $\chi_R$ are the bond dimensions between the subsystems $\mathcal{C}^\ell_n$ and $\bar{\mathcal{C}}^\ell_n$ on the left and right side, respectively.
It is therefore helpful to implement a cost model to predict which strategy will be most efficient in each case, taking into account the cost of contraction.
It is also advisable to store partially contracted tensors for later use with larger subsystems.
A small first-in-first-out cache is effective if the subsystem entropies are calculated in order of increasing subsystem size, and the cost model should account for the existence of cached objects.
Another approach for calculating entropies employs the MPS site-swap algorithm~\cite{Stoudenmire2010} to move bulk subsystems to the edge, such that double-cut entropies become single-cut entropies. 
Since singular values can be truncated following each swap, this approach may be suitable when an approximate information lattice is sufficient. 
We caution, however, that severe truncations risk breaking the strong subadditivity of subsystem entropies.
We emphasize that entries at the center of the information lattice have the highest computational cost. 
Such values, which are given by entropies of subsystems of scale $\ell \approx L/2$ that lie deep in the bulk, are dominant in ergodic states (as exemplified in Fig.~\ref{fig:DMRG}).
On the other hand, capturing both local features (dominant, for instance, in Anderson and many-body localized systems) and large-scale features (as topological edge modes) can always be done efficiently.

\bibliography{biblio}

\end{document}